\documentstyle[12pt,oldlfont,osa]{revtex}
\title{Method for Calculating One-Exciton Absorption Spectrum
or Space-Restricted Lattices}
\author{G.G.Kozlov}
\begin{document}
\maketitle
\begin{abstract}
The problem of the one-exciton absorption spectrum is considered
for the lattice of two-level interacting atoms whose initial
energy splitting depends on the coordinate. It is shown that for
some types of interatomic interaction, this problem can be
reduced to a differential equation of Schrodinger type, which,
in some cases, can be solved in closed form. By way of example,
problems on the one dimensional chain having an initial
splitting jump and on the spherical cluster are solved. In both
cases analitical solutions are obtained, which considerably
reduce the time required for calculations of one-dimensional
systems and permit the calculation of three-dimensional
spherical clusters, whereas the numerical calculation becomes
impossible for the clusters with a radius larger than ten
lattice constants. The spectra calculated for small systems
completely coincide with the spectra obtained by means of
numerical diagonalization. The proposed method may appear to be
convinient for solving problems related to Frenkel excitons in
systems having no translation symmetry (disordered systems).
\end{abstract}
\section{Introduction}

The recent advent of technology of manufacturing small-sized
crystalline structures has stimulated spectroscopic studies of
such objects. These objects include, for example, quantum wells,
quantum wires, J-aggregates and nanocrystals. Physical
properties of such objects (absorption spectra, the specific
heat, etc.) are significantly determined by their restricted
sizes. From the point of viev of optical spectroscopy, such
systems can be treated, in some cases, as fragments of crystal
lattice, and their absorption spectra can be interpreted using
the concept of excitons. In this case it is useful to have
simple computational technique for estimating the scale of
effects related to the restricted size of these systems. This
paper deals with one such method.

In the second section, the mathematical scheme of the method is
presented, and the differential equation is obtained for the
exciton Green's function required for calculation of the
absorption spectrum. In the third section, the one dimensional
exponential model of an exciton is formulated, in which the
interatomic interaction exponentially varies in space, and the
equation obtained in the second section has a simpler form and
can be exactly solved in some cases. As an example, expressions
are obtained for the absorption spectra of a finite chain and a
chain in which the distribution of the initial splitting energy
has the form of meander. In the forth section, the suggested
method is generalized to the case of three-dimensional
restricted systems. In this case, the equation for Green's
function simplified, if the interatomic interaction is discribed
by the Yukawa potential. In this section, the expression derived
for the absorption spectrum of a spherical cluster. The spectra
obtained in Sections 3 and 4 exhibit a well developed structure
and strongly differ from the monochromatic spectrum of an
infinite crystal. Almost all spectra were tested by means of a
direct computer diagonalization, and, in all cases, complete
agreement was achieved. The computer calculation of the spectra
of one-dimensional chains with lenths exceeding 100 lattice
constants $a$ takes much more time than the calculation
according the above expressions. Numerical computer-aided
analysis of the spectral structure of a three-dimensional
cluster requires several hours for the clusters radius $10a$. At
the same time, as shown in Section 4, the spectrum of a cluster
with radius of $30a$ exhibits the well-developed structure
related to the restricted size of the cluster. The expressions
obtained in the Section 4 can be used for estimating effects
caused by the restricted size of aspherical clusters.

Note that all the expressions obtained in this paper are
essentially based on the exponential (one-dimensional problems)
and Yukawa (three-dimensional problem) types of interatomic
interaction. This, however, does not appreciably restrict the
field of application of the above method for the following
reason. The exciton spectrum is often discribed in effective
mass approximation, which adequately reflects the dispersion law
only for long-wavelength excitons. The presence of two
parameters in exponential (Yukawa) potential allows one to
simulate not only the effective mass of an exciton but also the
exciton bandgap.

\section{ Equation for the exciton Green's function}

Consider a regular lattice of two-level atoms. We assume that
the initial (neglecting the interatomic interaction) splitting
of atomic levels is a given function $\varepsilon (r)$ of the
atomic coordinate $r$. Consider also the interatomic
interaction, which can cause the energy transfer between atoms,
the dependence of this interaction on the interatomic distance
being determined by the function $w(r)$ We will study only one
particle excited states of this system and the absorption
spectrum ralated to the transitions to these states from the
ground state, i.e. the one-exciton absorption spectrum. If the
transition dipole moment is the same for all atoms in the
lattice, this spectrum can be found by diagonalization of the
matrix $H_{rr'}=\delta_{rr'}\varepsilon (r)+w(r-r')$, the
positions of spectral lines being determined by eigenvalues, and
their intensites--by the squares of sums of components of
corresponding column eignvectors (it is in this way that squares
of moduli of corresponding matrix elements upon transitions from
ground state are calculated). Below, however, we will use the
following method of calculation of the above spectrum. We will
construct Green's function $g_{rr'}=(E-H)^{-1}_{rr'}$ for matrix
$H$ and define the function $G(r,E)$ as follows

$$G(r,E)=\sum\limits_l g_{rl}$$

Then, the absorption spectrum $A(E)$ , correct to insignificant
factors , has the form \cite{Onodera}
 
\begin{equation}
A(E)=-\hbox{Im}\sum\limits_r G(r,E+i\delta)
\end{equation}

each line in the spectrum being discribed by Lorentzian with the
half-width $\delta$. By considering $w(r-r')$ in the matrix $H$
as a perturbation, we write Dyson equation for Green's function

$$g_{rr'}=(E-\varepsilon (r))^{-1} \delta_{rr'} +
\sum\limits_l (E-\varepsilon(r))^{-1} w(r-l)g_{lr'}$$

Summation over $r'$ yields the equation for the function $G$ in
which we will pass from summation over $l$ to the integration:
 
\begin{equation}
G(r)(E-\varepsilon(r))=1+\rho\int w(r-r')G(r')dr'
\end{equation}

Here $\rho$ is the lattice density. The passage from summation
to integration in (2) is the only approximation we use, and it
should be noted that such passage is acceptable. The sum over
$l$ in the Dyson equation is close to the integral in (2),
provided the function $w(r-r')G(r')$ changes only slightly when
$r'$ changes by a lattice constant $a$. For this to happen,
first, it is required that $R>a$. Second, $G(r)$ also should not
strongly change on a scale $\sim a$, which should be checked
after calculations according to the procedure suggested below.
Note that in the region of absorption spectrum, the function $G$
oscillates at the spacial frequency of the resonant exciton.
Because the main contribution to the absorption spectrum (1) is
made by long-wavelength excitons, the passage from summation to
integration in (2) is justified. It should be emphasized that
the function $G$ in (1) describes the absorption spectrum rather
than the density of states whose calculation would require the
consideration of short-wavelenght excitons.

Let us now define the function $\Psi (r)\equiv
G(r)(E-\varepsilon(r))$ It follows from (2) that

\begin{equation}
\Psi(r)=\rho\int w(r-r')\bigg[\lambda^2 + {\Psi(r')\over
E-\varepsilon(r')}\bigg]dr'
\end{equation}

where
 
$$\lambda^{-2}\equiv\rho\int w(r)dr $$

Integral equation (3) can be reduced to the differential
equation in the following way. One can construct the
differential operator $L(\partial/\partial r)$ for wide class of
functions $w(r)$, so that $w(r)$ will be the source function
(Green's function) for this operator, i.e.,

\begin{equation}
L(\partial/\partial r)w(r)=\delta(r)
\end{equation}

To do this, equation (4) should be written in the Fourier
representation
 
$$w(r)={1\over 2\pi}\int e^{jkr} \tilde w(k)dk,$$

\begin{equation}
L(jk)\tilde w(k)=1
\end{equation}

The function $L$ is determined from the latter equation.
Consider now the inhomogeneous equation
 
\begin{equation}
L(\partial/\partial r)\Psi=\Phi(r)
\end{equation}

The particular solution of this equation can be obtained with
the help of the source function $w(r)$:
 
\begin{equation}
\Psi(r)=\int w(r-r')\Phi(r')dr'
\end{equation}

If

$$\Phi=\rho\bigg[\lambda^2+{\Psi(r)\over
E-\varepsilon(r)}\bigg],$$
 
then expression (7) coincides with (3) . Therefore, the required
solution $\Psi(r)$ of integral equation (3) satisfies the
differential equation
 
\begin{equation}
L(\partial/\partial r)\Psi(r)=\rho\bigg[\lambda^2+ {\Psi(r)\over
E-\varepsilon(r)}\bigg]
\end{equation}

In a number of cases, it is easier to solve the equation (8)
than (3). If the function $\Psi$ is known, the absorption
spectrum is detrmined by an expression that is similar to (1):
 
\begin{equation}
A(E)=-\hbox{Im}\int\limits_S{\Psi(r)\over
E+j\delta-\varepsilon(r)}dr
\end{equation}

where $S$ is lattice area.

\section{One-dimensional chain with exponential interatomic
interaction}

Consider a one-dimensional chain with interatomic interaction of
the form $w(r)=V\exp (-|r|/R)$. One can obtain from (5) the
following expression for the operator $L(\partial/\partial r)$
 
$$ L(d/dr)={R\over 2V}\bigg(R^{-2}-d^2/dr^2\bigg)$$

and equation (8) is transformed to the equation
\begin{equation}
\Psi^{\prime\prime} + \beta^2 \Psi =-R^{-2},
\end{equation}
$$\beta^2=R^{-2}\bigg({W\over E-\varepsilon(r)}-1\bigg) $$ $$
W=2\rho VR,$$

Which is similar to the inhomogeneous Schrodinger equation with
the 'potential' $\beta^2(r)$. For this reason, the function
$\Psi$ has a continuous first derivative. Let us find $\Psi$ for
a finite chain whose ends have coordinates $\pm l$. In this
case, $\varepsilon (r)$ has the form
 
$$
\varepsilon(r)=\cases{ \infty & $r\leq -l$ \cr 
\infty & $ r\geq l$ \cr
0 &$ -l < r < l$ } $$

i.e., the region outside the chain is as if filled by infinitely
detuned atoms to which the excitation cannot arrive. Of course,
this is only the 'variable' component of the splitting. The
'constant' component results in a shift of the spectrum as a
whole, and we will not take it into account below. In each
region , the solution of (10) can be obtained quite easily.
Taking into account the symmetry of the chain, the solution can
be written in the form
 
$$ \Psi(r)=\cases{C_1\cos(\beta r)-1/ (\beta R)^2 & $-l<r<l$\cr
C_2\exp(-r/R)+1 & $r>l$\cr C_2\exp(r/R)+1 & $r<-l$ } $$

where
 
$$ \beta^2=R^{-2}\bigg({W\over E} -1 \bigg)$$

If $E$ is not too close to zero (see below), then $\beta$ is the
wave number of an exciton with energy $E$. In regions outside
the chain, we retained only decreasing exponentials, because the
function $G$ vanishes there. This can be done passing to the
limit $\varepsilon \rightarrow \infty$ for $r<-l$ and $r>l$, if
the function $\Psi$ is limited. Constants $C_1$ and $C_2$ are
determined by the continuity condition for the function $\Psi$
and its derivative at the chains ends. By dropping simple
calculations, we present the final result for the function $G$
inside the chain:
 
\begin{equation}
G(r)={1\over E-W}\bigg(1+{W\over E}{\cos(\beta r)\over
\beta l \sin(\beta l)-\cos(\beta l)}\bigg) 
\end{equation}

Let us indicate the region of energies $E$ where the absorption
spectrum calculated from (1) and (11) can be used. The passage
from summation to integration in Dyson equation (2) will be
justified, if the spacial half-period of $G$ (equal to
$\pi/\beta$) is grater than the lattice constant $a$. This
results in the condition

\begin{equation}
E>E_C\equiv {W\over 1+(\pi R/a)^2}
\end{equation}

Figure 1 (There are no figures in this version, see figures in
Optics and Spectroscopy, Vol.82, No. 2, 1997, pp 242-246) shows
the absorption spectrum calculated from (11) and (1) for the
chain parameters $l=50,R=2,V=1,$ and $\delta=0.1$, the energy
$E$ being plotted in units of $V$. The vertical bar along the
abscissa axis shows the energy $E_C$. One can see that in the
"forbidden" region, the spectrum is in fact absent, and this
region is of no interest. The spectrum of this chain was also
obtained by numerical diagonalization and proved to be
completely coincident with the spectrum in Fig.1.

Let us also consider here the somewhat less trivial case of a
chain in which $\varepsilon(r)$ has the form of a meander:
 
$$ \varepsilon(r)=\cases {\infty & $r<-h$ or $r>h$\cr
\varepsilon & $l<r<h$ or $-l>r>-h$\cr
0 & $-l<r<l$\cr} $$

We assume for the simplicity of calculations, that this chain is
closed to a ring, i.e., the point $h$ coincides with $-h$. The
solution of (10) is obtained in similar manner, the only
difference being that it is necessary to require the values of
the function $\Psi$ and its derivative to be the same in the
points $h$ and $-h$. The result for integrated Green's function,
whose imaginary part determines the absorption spectrum, is

\begin{equation}
{1\over 2}\int\limits_{-h}\limits^h G dr={l\over E-W}+ {h-l\over
E-W-\varepsilon}+{\beta_1^2-\beta_2^2\over \beta_1\beta_2}
{(W-E)^{-1}-(W+\varepsilon-E)^{-1}\over
\beta_2\hbox{ctg}(\beta_1l)+\beta_1
\hbox{ctg}(\beta_2(h-l))}
\end{equation}

where
 
$$ \beta_i= R^{-2}\bigg({W\over E-\varepsilon_i}-1\bigg)$$ $$
\varepsilon_{1,2}=0,\varepsilon $$

In this case, condition (12) should be supplemented by the
requirement
 
\begin{equation}
\varepsilon- E_C>E>\varepsilon+E_C
\end{equation}

because for energies lying within the $E_C$ vicinity of
$\varepsilon$, the function $G$ has a spatial period smaller
than $a$ on the chain fragments with initial splitting
$\varepsilon$. Note, however, that in this case, the
contribution from the above fragments will be small compared to
that from the rest of the chain, where the function $G$
oscillates not so rapidly. For this reason, the absorption
spectrum can be realistic even when condition (4) is violated.

The spectrum of the chain with a jump of the initial splitting
is presented in Fig. 2. The parameters are:
$h=30,l=10,R=2,V=1,\delta=0.1,$ and $\varepsilon=2.5$; the
energy is plotted in units of $V$. The spectrum in Fig. 2a is
obtained by numerical diagonalisation, and that in Fig 2b is
calculated from (13). The regions of forbidden energies are
indecated by a vertical bar $E_C$ and horizontal bar [condition
(14)]. For the chains of lenght $\sim 100a$, calculation by
means of the above equations takes much less time than numerical
diagonalization. In the next section the problem of
three-demensional spherical cluster will be solved. The
numerical calculation of such a claster with radius $>20a$
represents a vast computer problem, while clusters with radii
$\sim 150a$ still exhibit the distinct structure, related to
their finite size.

\section{Spherical cluster with interatomic interaction of
Yukawa type}

The approach presented in section 2 can be generalized to the
case of an arbitrary demensionality. In this section we consider
a three-dimensional lattice with an inteaction of $w(r)=V \exp
(-r/R)/r$ (the Yukawa potential). It is well known [and can be
easily verified from (5)] that
 
$$\bigg(\Delta-R^{-2}\bigg)\bigg({1\over
4\pi}{\exp(-|r|/R)\over|r|}\bigg)=
\delta(r)$$

After calculating the integral $\lambda^{-2}$ in (3) in the
spherical coordinate system, we find that in this case, equation
(8) again has the form of the inhomogeneous Shredinger equation
 
\begin{equation}
\Delta\Psi+R^{-2}\bigg({W\over E-\varepsilon(r)}-1\bigg)\Psi=-R^{-2}
\end{equation}

where
 
$$ W\equiv4\pi\rho V R^{2} $$

and, hence, function $\Psi$ is continuous with its first
derivative.

Let us now solve the problem on a spherical cluster with radius
$\sigma$, i.e., the function $\varepsilon (r)$ will be taken in
the form

$$
\varepsilon(r)= \cases{0 & $|r|<\sigma$\cr
\infty & $|r|>\sigma$}
$$

By representing the Laplace operator in (15) in spherical
coordinates, for function $\Psi$ (it has only the radial
dependence), we obtain the equation
 
\begin{equation}
\bigg({1\over r^2}{d\over dr}r^2{d\over dr}+\beta^2\bigg)\Psi=-{1\over
R^2}
\end{equation}

where

$$\beta^2=\bigg({W\over E-\varepsilon(r)}-1\bigg)R^{-2}$$

In our case,
 
$$
\beta^2=\cases{-R^{-2} & $r>\sigma$\cr
\big(W/E-1\big)R^{-2} & $r<\sigma$}
$$

The general solution of equation (16) for both regions has the
form

$$
\Psi=-\bigg( {1\over R\beta}\bigg)^2+a{\sin\beta r\over r}+b{\cos \beta
r
\over r}
$$

Taking into account the boundedness of $\Psi$ in the region
$r>\sigma$, we obtain that in this region,
  
\begin{equation}
\Psi=1+a_1{\exp(-r/R)\over r}
\end{equation}

Function should be also bounded at $r=0$. Taking this into
account, we obtain that in the region $r<\sigma$,
  
\begin{equation}
\Psi=EG_\infty+a_2{\sin\beta r\over r}
\end{equation}

where
 
$$ G_\infty\equiv(E-W)^{-1} $$

Function $EG_\infty$ is the solution of (15) in the case of an
infinite homogeneous lattice. For this reason, Im$G_\infty
(E+j\delta)$ determines the absorption spectrum of such a
lattice. Constants $a_1$ and $a_2$ are determined from
continuity conditions for the function $\Psi$ and its first
derivative for $r=\sigma$ The absorption spectrum is determined
from (9)
 
$$ A(E)=-\hbox{Im}\int\limits_0\limits^\sigma
{\Psi(r,E+j\delta)\over E+j\delta}r^2dr $$

which yields the result
 
\begin{equation}
A(E)=-\hbox{Im}\biggl\lbrace
\bigg(
{\sigma^3\over
3}-{W(R+\sigma)[\sin\sigma\alpha-\sigma\alpha\cos\sigma\alpha]
\over\alpha^2 (E+j\delta)(R\alpha\cos\sigma\alpha+\sin\sigma\alpha)}
\bigg)G_\infty(E+j\delta)
\biggr\rbrace
\end{equation}

where
 
$$
\alpha\equiv-{1\over R}\root 2 \of{{W\over E+j\delta}-1}
$$

Expression (19) contains two contributions: the bulk
contribution (the first term in parentheses) and the surface
contribution (the second term). The first term gives the
spectral line that corresponds to an infinite lattice, while the
second one is related to the structure caused by the finite size
of the cluster.

Figure 3 shows the absorption spectrum of a spherical cluster
calculated from (19) for the parameters $V=1,R=1, \sigma=5,$ and
$\delta=0.15$; the energy is plotted in units of $V$. The
vertical bar shows the energy $E_C$. The spectrum obtained by
numerical diagonalization coincides completely with the spectrum
in Fig. 3. However, it should be noted that the Yukawa potetial
has a singularity at $r=0$. For this reason, the question arises
upon numerical calculations as to what form the diagonal
elements of the $H_{rr'}$ matrix should take (see Section 2).
Although these matrix elements do not affect the shape of the
absorption spectrum, they may cause its shift. In the numerical
verification of the spectrum in Fig. 3, we assumed that
$H_{rr'}=0$, because in the calculation of intergrals with
Yukawa potential, the contribution from the region $r\sim 0$ is
insignificant due to the smallness of the volume $r^2 dr$, and,
in fact, the "effective potential" $w(r)r^2$ is used evrywhere,
which vanishes for $r=0$. Computer calculations performed with
other values of parameters are also in complete agreement with
(19).

Figure 4 shows the spectrum of spherical cluster for
$V=1,R=2,\sigma=30,$ and $\delta=0.6$. Comparison with numerical
diagonalization was not performed, because, in this case,
computer calculations are time consuming. Figure 4 shows that
despite large radius of the cluster, its spectrum differs from
that of infinite lattice (the singlet at $E=W$) and exhibits a
structure. Clusters with a radius of $150a$ also show a
noticeable structure (on the scale of the bandgap $W$).

This work was supported by the Russian Foundation for Basic
Research, project no. 93-02-16240.

\end{document}